\documentclass[prb,aps,10pt,showpacs,twocolumn,superscriptaddress]{revtex4}
\usepackage{graphicx,amssymb,amsmath,color}
\newcommand{\e}{\varepsilon}
\newcommand{\Nav}{M}
\newcommand{\Vdc}{V_\mathrm{dc}}
\newcommand{\lb}{\langle\!\langle}
\newcommand{\rb}{\rangle\!\rangle}
\begin{document}
\title{Counting Statistics and Dephasing Transition in an Electronic Mach-Zehnder Interferometer}
\author{A. Helzel }
\author{L. V. Litvin }
\affiliation{\mbox{Institut f\"{u}r experimentelle und angewandte Physik, Universit\"{a}t  Regensburg, D-93040 Regensburg, Germany}}
\author{I. P. Levkivskyi}
\affiliation{Theoretische Physik, ETH Zurich, CH-8093 Zurich, Switzerland}
\affiliation{Bogolyubov Institute for Theoretical Physics, 03680 Kiev, Ukraine}
\author{E. V. Sukhorukov}
\affiliation{\mbox{D\'{e}partement de Physique Th\'{e}orique, Universit\'{e} de Gen\`{e}ve, CH-1211 Gen\`{e}ve 4, Switzerland}}
\author{W. Wegscheider }
\affiliation{\mbox{Laboratorium f\"{u}r Festk\"orperphysik, HPF E 7, ETH Z\"urich, 8093 Z\"urich, Switzerland}}
\author{C. Strunk}
\affiliation{\mbox{Institut f\"{u}r experimentelle und angewandte Physik, Universit\"{a}t  Regensburg, D-93040 Regensburg, Germany}}
\date{\today}
\begin{abstract}
It was recently suggested that a novel type of phase transition may occur in the visibility of  electronic Mach-Zehnder Interferometers. Here, we present experimental evidence for the existence of this transition. The transition is induced by strongly non-Gaussian noise that originates from the strong coupling of a quantum point contact to the interferometer. We provide a transparent physical picture of the effect, by exploiting a close analogy to the neutrino-oscillations of particle physics. In addition, our experiment constitutes a probe of the singularity of the elusive full counting statistics of a quantum point contact.
\end{abstract}
%
%
\maketitle

\begin{figure*}[t]
\includegraphics[width=170mm]{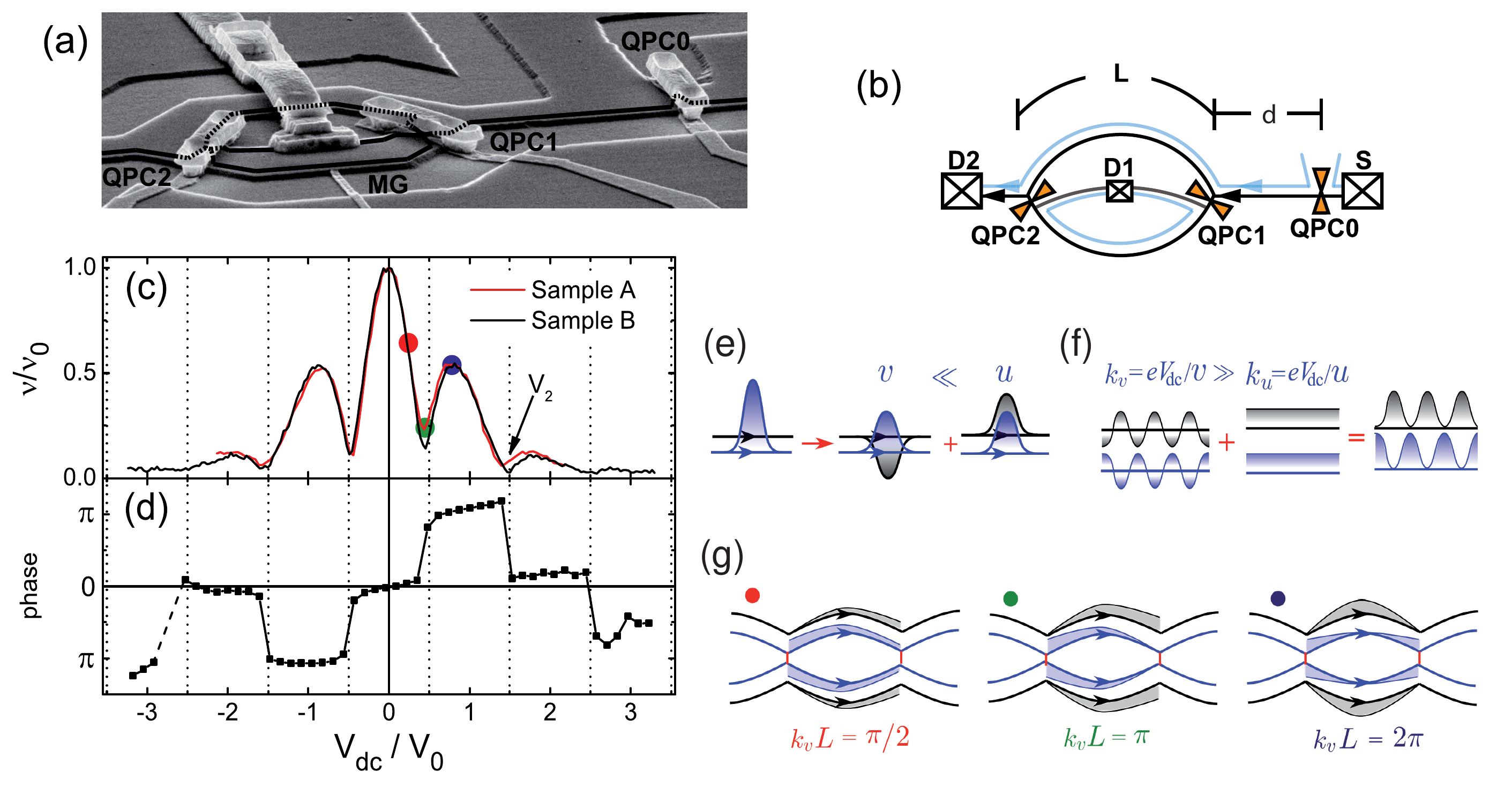}
\caption{\textbf{(a) Schematic} of the edge channels in the MZI connected to source and drain. The black line is the outer edge channel carrying interfering electrons and the light blue line represents the inner edge channel. \textbf{(b) Scanning electron micrograph} of the sample with marked gates QPC~0, QPC~1, QPC~2 and the modulation gate MG. The black line represents the edge channel used for the interference. QPCs~1 and 2 are set to half transmission and QPC~0 is set to transmit the outer edge with probability ${\cal T}_0$. Ramping the modulation gate alters the area between the electron paths and thus the Aharonov-Bohm phase. \textbf{(c) Multiple lobe structure:}  If a dc voltage is applied to the interfering edge channel [black line in panels (a, b)] the differential visibility shows a multiple lobe structure with several nodes, which is displayed for two samples, A (red line) and B (black line). When the visibility is scaled with respect to the zero bias maximum visibility $\nu_0$ and the voltage with respect to the node spacing $V_0$ the curves of the two samples collapse. \textbf{(d)} \textbf{Phase evolution:} The lobes appear as phase-rigid plateaus within the lobes and phase jumps of $\pi$ at the nodes. 
 \textbf{(e) Representation of an electron in terms of charge and dipole plasmon wave packets:} In the presence of strong Coulomb interactions, the edge eigenmodes are the fast charged plasmons and the slow dipole plasmons with the linear spectrum in the low-energy limit. \textbf{(f) Superposition of delocalized charge and dipole modes:} An electron injected at the energy $eV_{\rm dc}$ excites dipole plasmons with wavelength $k_v = eV_{\rm dc}/v$ and charged plasmons with wavelength $k_u = eV_{\rm dc}/u$, so that $k_u \ll k_v$. Such superposition of plasmons has alternating zeros and maxima along the edge. \textbf{(g) Secondary interference of collective modes:} At particular values of the voltage bias applied to the interferometer, the plasmon amplitude has a maximum or a minimum at the position of the second QPC, corresponding to constructive or destructive secondary interference.
}\label{char_lobes}
\end{figure*}

The recent discovery of a lobe-type behavior in the visibility of Aharonov-Bohm
oscillations in electronic Mach-Zehnder interferometers (MZI) has triggered extensive theoretical
studies in this field. These interferometers were implemented in the edge channels in the integer quantum Hall effect (QHE), mostly for filling factor {\it ff}$\,=2$ (see Fig.~\ref{char_lobes}a,b), \cite{heiblum1, roche, Leonid1}. 
Many sophisticated theories have been proposed to explain the side lobes.\cite{cheianov, chalker, sukh1,ginossar, sim1, chalker2, mirlin, mirlin2} Here we show that  this puzzling effect can be explained in a rather simple way, if two interacting edge channels are present. The underlying phenomenon turns out to be very similar to that of neutrino oscillations
in high-energy physics: neutrinos oscillate between different flavor states because they are created in a flavor eigenstate, which is not an eigenstate of the Hamiltonian. Similarly, when an electron wave packet is partitioned by a beam splitter, it excites a collective charge mode, which is not an eigenstate of our model Hamiltonian.\cite{sukh1}
At the second beam splitter of the interferometer, this leads to a secondary interference between the collective modes as a function of the applied voltage bias. This model can explain many of the experimental observations\cite{heiblum3, Leonid2, roche4, Basel, huynh}, most importantly the visibility lobes\cite{heiblum2} and the phase rigidity of the visibility.\cite{heiblum1,roche}

 Dephasing of the Aharanov-Bohm interference results from random fluctuations of the phase that are averaged out by the detector. In our devices  fluctuations are generated by an additional quantum point contact (QPC-0) in front of the interferometer input, which can be controlled by its transmission probability ${\cal T}_0$.   The noisy input current leads to charge fluctuations in the interferometer. The accumulated charge shifts the edge, which leads to the Aharonov-Bohm phase shift. Hence, the strong Coulomb interaction between the edge channels guarantees a strong coupling of the electrons in the interferometer to the noise.  The visibility will thus be suppressed by the charge fluctuations induced by partitioning of electrons at QPC-0. Most interestingly, the lobe-pattern was predicted to undergo a sudden change at ${\cal T}_0=1/2$ under such conditions.\cite{sukh2} This noise-induced transition provides an unique experimental signature of the non-Gaussian character of the noise in the MZI visibility.

Noise at zero frequency is described by the set of irreducible moments (cumulants) $\lb I^j\rb$, where $j=1, 2,\ldots$, of the currents' distribution. 
  Alternatively, the probability distribution can be described by the generator
$h(\lambda)=\sum_j\lb I^j\rb (i\lambda)^j/j!$ of
its full counting statistics (FCS).\cite{levitov}
For Gaussian noise the FCS generator contains only the first two terms
$h(\lambda)=\langle I\rangle i\lambda -\lb I^2\rb\lambda^2/2$, where $\langle I\rangle$ is the average current and $\lb I^2\rb$ is the zero frequency noise power.
The third cumulant $\lb I^3\rb$ vanishes at equilibrium, and thus it provides a measure of the deviations from Gaussian statistics in non-equilibrium state.\cite{LevitovReznikov, reulet, reznikov,T,H,ensslin_qdots, haug}

In noise detection schemes in mesoscopic physics the variable $\lambda$ plays the role of a coupling constant, connecting the noise source and the detector.\cite{levitov,sukh_averin} The coupling constant is typically very small, which demands long-time measurements.
In this case, the high-order cumulants are suppressed, because many random events contribute
to the detector output signal, and by virtue of the central limit theorem the noise becomes effectively Gaussian.
Signatures of non-Gaussian noise have presumably been observed via
its effect on the visibility of Aharonov-Bohm oscillations in MZIs.\cite{heiblum4}
Introducing noise into the co-propagating inner edge channel, which was electrically disconnected from the interfering
outer edge channel, was shown to reduce the visibility. A particular V-shaped dependence of the visibility $\nu$ of
the interference on the transmission of the co-propagating channel was observed and attributed to the non-Gaussian
character of the noise. However, that observation could not be reproduced in a similar experiment,\cite{roche3}
nor was it linked to the peculiar multiple lobe structure\cite{heiblum2,Leonid2,huynh} in $\nu(\Vdc)$ of such interferometers.

Here, we present experimental evidence for a noise
induced non-equilibrium phase transition that was predicted to occur at ${\cal T}_0=1/2$.\cite{sukh2}
This novel type of non-equilibrium phase transition is caused by a singularity in the FCS generator $h(\lambda)$
of the QPC occurring at $\lambda=\pi$, which leads to a singular dephasing rate of
the interferometer under non-equilibrium conditions. This effect can be linked to the perfect entanglement
between the MZI and the QPC, \cite{sukh_averin} as a consequence of the strong coupling between the two,
and may be seen as a manifestation of the quantum origin of shot noise
(for details, see the Sec.\ VI of the supplement).

We start by addressing the physics of multiple side lobes in the visibility of AB oscillations in MZIs.
This phenomenon was observed experimentally in the visibility function
$\nu(V_{\textrm{dc}})=(G_\mathrm{max}-G_\mathrm{min})/(G_\mathrm{max}+G_\mathrm{min})$, where
$G$ is the differential conductance of the interferometer,
for a noiseless incident beam and filling factors $2>\emph{ff}>1.5$.\cite{heiblum2,Leonid2,huynh}
Fig.~\ref{char_lobes} shows a schematic (a) and an electron micrograph (b) of a typical sample, the visibility $\nu$ normalized to its value $\nu_0$ at $\Vdc=0$ (c), and the corresponding phase evolution (d). To explain these side lobes, several theories have been proposed.\cite{cheianov,chalker,ginossar,sim1,sukh1,chalker2,mirlin, mirlin2} Most of the experimental observations can be consistently explained by modelling the QHE edge states as chiral quantum wires supporting co-propagating magneto-plasmon modes at \emph{ff}$=2$ (for details, see the Sec.\ IV of the supplement.). Coulomb interaction between these modes results in a pair of charged and neutral (dipolar) plasmon modes propagating at different velocities along the edge (see Fig.~\ref{char_lobes}e). Because of this velocity difference charge oscillations introduced in one of the edges can be transferred from one edge to the other, resulting in a collapse of the interference pattern at certain equidistant voltages $V_k=(k-1/2)V_0$, where the node spacing $V_0$ is inverse proportional to the arms length $L$.\cite{sukh1}
This effect is remarkably similar to neutrino oscillations, \cite{n-osc-first} a phenomenon in high-energy physics that manifests itself
in the periodic change of the flavor quantum number of freely propagating neutrinos. \cite{neutrino}

Next, we sketch the main idea of the theory in Ref.~\onlinecite{sukh1}, while further details are presented in the Sec.~V of the supplement.
The many-body wave function of the two interacting edge channels in one of the arms of the interferometer
can be decomposed in two components: the charge part $|N\rangle$, where
$N$ is total number of electrons in one edge of the interferometer, and the plasmon part $|\mbox{plasmons}\rangle$. An electron tunneling at the first beam splitter
(QPC~1) changes the number $N$ by one and excites plasmons: $|N\rangle\to|\psi_{N+1}\rangle=|N+1\rangle|\mbox{plasmons}\rangle$ .
Two components evolve to the final state $|\psi'_{N+1}\rangle$ at  the second beam splitter (QPC~2) and acquire the specific phase shifts.
 Combining the two components of the wave functions $|\psi_{N+1}\rangle$ and $|\psi'_{N+1}\rangle$
results in a total probability amplitude for the detection of charge in one of the contacts, say D2:
\begin{equation}\label{eq:overlap}
\langle \psi_{N+1}|\psi_{N+1}^\prime\rangle=e^{-i\pi \Nav}\cdot\frac{1+e^{i2\pi \Nav}}{2}=\cos\left(\pi \Nav\right),
\end{equation}
where the dimensionless parameter
\begin{equation}\label{eq:N_vs_epsilon_L}
\Nav\ =\  \frac{e\Vdc L}{2\pi\hbar v}.
\end{equation}
 can be interpreted as the average number of {\it excess} electrons injected by the source into the outer edge channel during the time $\tau=L/v$ taken by the slow dipole mode to reach the QPC~2 (propagating at the speed $v$).
The first factor in Eq.~(\ref{eq:overlap}) is the phase shift resulting from the voltage induced accumulation of $\Nav/2$ electrons in the outer edge channel, while the second comes from the interference of charge and dipolar plasmons.

The result (\ref{eq:overlap}) and (\ref{eq:N_vs_epsilon_L}) indicates that the lobe pattern observed in Fig.~\ref{char_lobes}d
originates from the secondary interference controlled by $\Nav$. The phase rigidity follows from the fact that the overlap is a real function.
The only free parameter in the theory of Ref.~\onlinecite{sukh1} is the velocity $v$ of dipole plasmons. It is reflected in a characteristic energy
\begin{equation}
\label{EL}
\e_L=eV_0=2\pi\hbar v/L
\end{equation}
that depends on $L$ only.\cite{footnote_epsilon_L}
For two interferometers of different arm length, $\e_L$ is extracted from the position of the $2^\mathrm{nd}$ visibility node at voltage $V_2$ in Fig.~\ref{char_lobes}c, which corresponds to a value of $V_0\cdot L=2.8\cdot 10^{-10}\,\mathrm{Vm}$, or $v=1.1\cdot10^5\;$m/s. For sample~A (B) we obtain $\e_L=45.8\,(30.6)\,\mu$eV.
In Fig.~\ref{char_lobes}c a strong damping of the oscillations is seen that has been phenomenologically accounted for by an additional factor $D(\Vdc)=\exp[-(e\Vdc)^2/2\e_0^2]$  in Ref.~\onlinecite{Leonid2}  with a characteristic energy $\e_0$ that is of the same order of magnitude as $\e_L$. The damping is an effect of inelastic scattering,\cite{altimiras} which is not contained in the theory. In the following the damping factor will be removed by normalizing the visibility $\nu(\Vdc,{\cal T}_0)$ with respect to  $\nu(\Vdc,{\cal T}_0=1)$.

\begin{figure}[t]
\includegraphics[width=85mm]{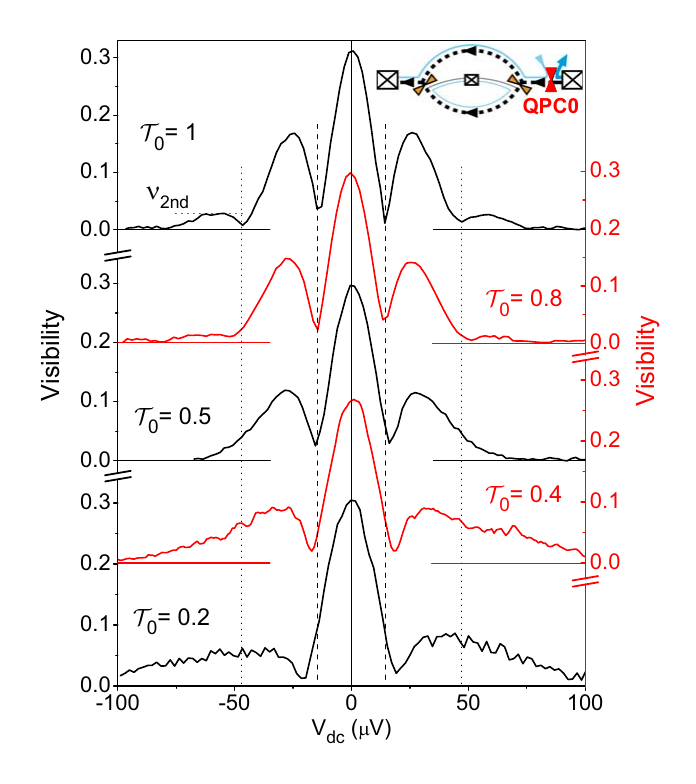}
\caption{\textbf{Visibility lobe structure of sample B ($L=8.7\,\mu$m) for different ${\cal T}_0$:} When closing QPC~0 (see inset) below ${\cal T}_0=1$ the node positions remain essentially unchanged for transmissions ${\cal T}_0\geq 0.5$, i.e., multiple side lobes with the same widths of lobes and position of nodes as for ${\cal T}_0=1$. The lobe height is strongly reduced at finite dc bias. Below ${\cal T}_0=1/2$ the lobe structure changes from the multiple side lobe behavior to a single side lobe behavior. The central lobe width increases with decreasing transmission and the first side lobes are stretched to large bias. }
\label{lobechange_new}
\end{figure}

Next, we give an elementary argument of how the MZI visibility is connected to the FCS generator $h(\lambda)$
of the quantum point contact QPC~0 with the transparency ${\cal T}_0$ at the MZI input (see Fig.~\ref{char_lobes}),
which is known to take the form:
\cite{levitov}
\begin{equation}\label{FCS}
h(\lambda)=\frac{e V_{\rm dc}}{2\pi\hbar}\ln[1+{\cal T}_0(e^{i\lambda}-1)].
\end{equation}
Note, that this function displays for $\lambda=\pm\pi$ a singularity at transmission ${\cal T}_0=1/2$,
which in the context of quantum measurements reflects perfect entanglement between the system (MZI) and
the detector (QPC) as a consequence of the strong coupling between the two.\cite{sukh_averin}
The essential point of Ref.~\onlinecite{sukh2} is that the overlap of interfering quantum states (Eq.~\ref{eq:overlap}) depends on the number $m$ of excess electrons in the interferometer, injected during the time interval $\tau=L/v$. If the transparency of QPC~0 ${\cal T}_0=1$ we have $m=\Nav$ (see Eq.~\ref{eq:N_vs_epsilon_L}). Upon reducing ${\cal T}_0$ below one $m<\Nav$ becomes a random variable,
which fluctuates according to the binomial statistics between $0$
and $\Nav$ with the average value $\langle m \rangle={\cal T}_0\Nav$.

\begin{figure}
\includegraphics[width=85mm]{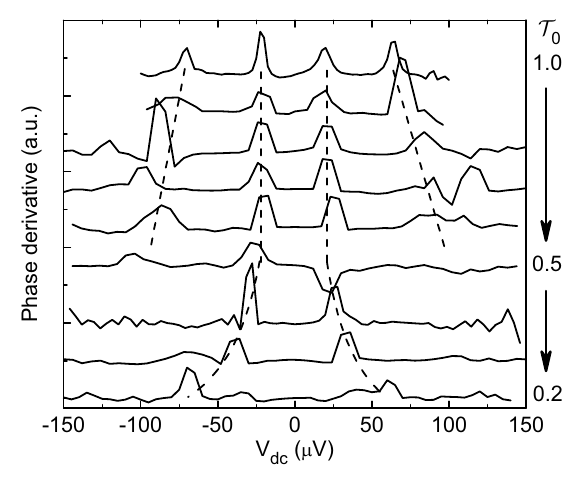}
\caption{\textbf{Phase changes of sample A ($L=6.5\,\mu\mathrm{m}$) for different ${\cal T}_0$:} Similar to Fig.~\ref{char_lobes}d the steps in the phase evolution at visibility nodes become discernible as peaks in the numerical derivative $d\varphi(\Vdc)/d\Vdc$ of the phase for different transmission of QPC~0 (from 1.0 to 0.2 in steps of 0.1). The width of the inner pair of phase steps remains nearly fixed for ${\cal T}_0>0.5$. Upon a further decrease of ${\cal T}_0$  the width of the central lobe is increasing. }
\label{lobesphase}
\end{figure}

For large $\Nav\gg1$, we can approximate $m$ to be integer. To
 find the visibility of the interference pattern $\nu$
one needs to average the overlap $\langle\psi_{N+1}|\psi'_{N+1}\rangle$
over the possible particle numbers:
\begin{equation}\label{eq:visibility}
\nu(\Nav)=\left|\sum_{m=0}^\Nav P(\Nav,m)\cos(\pi m)\right|,
\end{equation}
where
$P(\Nav,m)=B_\Nav^m{\cal T}_0^m(1-{\cal T}_0)^{\Nav-m}$ is the
 the binomial distribution of electron transmissions at the QPC~0,
 and $B_\Nav^m$ are the binomial coefficients.
The visibility (\ref{eq:visibility}) is thus connected to the FCS generator
(\ref{FCS}) via the Fourier transform $\sum_{m=0}^\Nav P(\Nav,m)e^{i\lambda m}=\exp[\tau h(\lambda)]$. 
By setting $\lambda=\pi$, one obtains
\begin{equation}\label{Eq:average_nu}
\nu(\Nav)=\big|\cos[\Nav\cdot\Theta(2{\cal T}_0-1)]\big|\cdot\exp\big[\Nav\cdot\ln\left|2{\cal T}_0-1\right|\big],
\end{equation}
where $\Theta$ is the Heaviside step function, and the non-analytic structure of $\nu$ stems from the singularity in $h(\lambda)$ at $\lambda=\pi$ and ${\cal T}_0=1/2$.
This is the origin of the noise-induced phase transition predicted in Ref.~\onlinecite{sukh2}.

The result in Eq.~\ref{Eq:average_nu} contains an oscillatory and a dephasing factor, which both depend on $\Nav$ and ${\cal T}_0$. For ${\cal T}_0>1/2$ multiple side lobes are expected, with visibility nodes that remain independent of $\Vdc$.  For ${\cal T}_0<1$ the normalized visibility is exponentially damped at large $\Vdc$. From the logarithm in the dephasing term in Eq.~\ref{Eq:average_nu} one expects a divergence of the dephasing rate at ${\cal T}_0=1/2$. The more rigorous treatment in Ref.~\onlinecite{sukh2} and our numerical calculations (see supplementary material) take into account the quantum fluctuations of $m$ and result in a divergence of the spacing between visibility nodes, when ${\cal T}_0$ approaches 1/2.

\begin{figure*}[t]
\includegraphics[width=165mm]{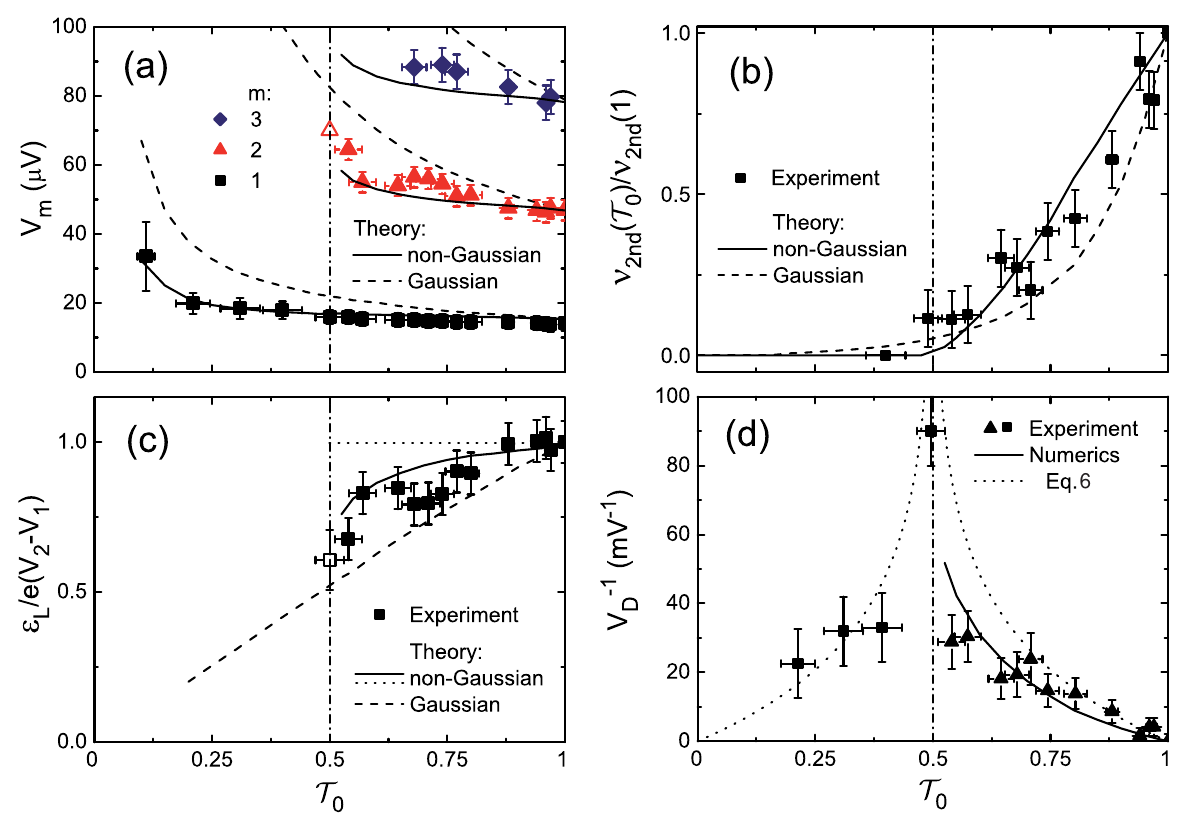}
\caption{\textbf{Nodes, lobes, and the phase transition: (a)}
Dots: Measured visibility node positions $V_m$ vs.~the transparency of QPC~0. The nodes shift outward, and disappear for $m>1$ at ${\cal T}_0=1/2$. Solid lines: Predictions of the full theory using the characteristic energy $\e_L=30.6\,\mu$eV extracted from Fig.~\ref{char_lobes}c. Dashed lines: Gaussian approximation of the theory.
\textbf{(b)} Dots: Maximal visibility  ($\nu_{2^\mathrm{nd}}$) of the second side lobes [see Fig.~\ref{lobechange_new} (sample B)], normalized with respect to $\nu_{2^\mathrm{nd}}({\cal T}_0=1)$. Solid (dashed) line: theoretical prediction for the full theory (its Gaussian approximation). While the Gaussian approximation predicted multiple side lobes for all values of ${\cal T}_0$, they appear in the full theory only for ${\cal T}_0=0.5$.
\textbf{(c)} Inverse normalized node spacing $e(V_2-V_1)/\e_L$ as a signature of the non-equilibrium phase transition.  Full line: full numerical calculation of the node spacing. Dotted line: expectation from the approximation of Eq.~\ref{eq:N_vs_epsilon_L}. Dashed line: Gaussian approximation of the full theory. The dash-dotted line indicates the point of the phase transition. In the presence of non-Gaussian noise the inverse node spacing vanished abruptly at the transition point ${\cal T}_0=1/2$.
\textbf{(d)} Dephasing rate $V^{-1}_D$ extracted from the exponential decay of $\nu(\Vdc)$ in Fig.~\ref{lobechange_new} for ${\cal T}_0\leq1/2$ (squares) and the amplitude of the $\nu_{2^\mathrm{nd}}(\Vdc)$ in (b) for ${\cal T}_0>1/2$  (triangles). The dotted line represents the dephasing rate extracted from Eq.~\ref{Eq:average_nu}, while the solid line is the result of the numerical simulations. No free parameters are adjusted in the theory curves of all four panels.
}
\label{lobezeros_nu2nd}
\end{figure*}

In the following, we show that all of these rich and intriguing predictions are observable in our experiment, and provide clear evidence for the non-Gaussian character of the noise.
In Fig.~\ref{lobechange_new} we present the evolution of the lobe pattern when we introduce noise to the interfering edge channel of the MZI by closing QPC~0 (see the inset in Fig.~\ref{lobechange_new}) so that the outer edge channel is only  partially transmitted. For ${\cal T}_0\geq0.5$ the number and position of the visibility nodes, stays (almost) constant, while the height of the second side lobe is gradually suppressed. The number of the side lobes and the phase rigidity is more clearly seen in the derivative $d\varphi(\Vdc)/d\Vdc$ of the phase plotted in Fig.~\ref{lobesphase}. Close to ${\cal T}_0\simeq0.5$ a second node is hard to see in Fig.~\ref{lobechange_new}, but a drop to zero visibility and the phase jump is still apparent in the gate modulation of the MZI-current in Fig.~\ref{lobesphase}.  Some structure is also seen at ${\cal T}_0=0.5$  (see supplementary material).

The situation changes drastically for ${\cal T}_0\leq0.5$ when multiple nodes and multiple side lobes disappear abruptly. This is seen very clearly in Fig.~\ref{lobesphase}. The central lobe width is increasing with decreasing QPC~0 transmission, i.e., the remaining single phase step shifts to high voltages and no additional nodes can be seen (see Fig.~\ref{lobechange_new}). 

Next we present a detailed and quantitative comparison of the dependence of the measured position of the visibility nodes and the height of the second side lobe on ${\cal T}_0$ with numerical calculations (see supplementary material) that extend the predictions of Ref.~\onlinecite{sukh2}. In Fig.~\ref{lobezeros_nu2nd}a we plot the positions $V_\mathrm{m}$ of the first three visibility nodes vs.~${\cal T}_0$ for sample~B. For ${\cal T}_0=1$ these are expected at $V_k=(k-1/2)\,\e_L/e$. Only the first visibility node is observed in the whole range of transparencies $0.1\leq{\cal T}_0\leq1$. The presence of the $2^{\mathrm{nd}}$ and $3^{\mathrm{rd}}$ visibility node is limited to the range $0.5\leq{\cal T}_0\leq1$, as predicted for non-Gaussian noise only. For decreasing transparency ${\cal T}_0$ the nodes shift outward in a way that is captured quantitatively by our numerical calculations (full lines) when using the value for the velocity of dipolar plasmons determined from Fig.~\ref{char_lobes}c. The dashed lines arise, when Gaussian noise is modelled by truncating the Taylor expansion of the FCS-generator (Eq.~\ref{FCS}) in $\lambda$ after the quadratic term. In this case, a stronger variation of all visibility nodes with ${\cal T}_0$ is expected, and the $2^{\mathrm{nd}}$ and $3^{\mathrm{rd}}$ node should exist also below ${\cal T}_0=0.5$. The observed absence of visibility nodes with $k>1$ is thus a strong evidence for the non-Gaussian character of the noise. The position of the $1^{\mathrm{st}}$ visibility node is slightly reduced with respect to the expected value $\e_L/2e$.

According to our theory, the amplitude of the second side lobe $\nu_{\textrm{2nd}}({\cal T}_0)$ vanishes in a universal quasi-linear fashion near ${\cal T}_0=0.5$, i.e., it is independent of the system parameters $v$ and $L$. In Fig.~\ref{lobezeros_nu2nd}b we present the ratio $\nu_{\textrm{2nd}}({\cal T}_0)/\nu_{\textrm{2nd}}({\cal T}_0=1)$ of the maximal visibility amplitudes $\nu_{\textrm{2nd}}({\cal T}_0)$ determined form sweeps of the modulation gate. The normalization with respect to $\nu_{\textrm{2nd}}({\cal T}_0=1)$ is necessary, to divide out the additional dephasing factor $D(\Vdc)$ discussed above.  Within the limits of the experimental accuracy, our data are again in good agreement with the theory.

In order to illustrate the character of the transition, and demonstrate the agreement of the measured data with the structure of Eq.~\ref{Eq:average_nu} we illustrate the peculiar variation of the arguments of the $\cos$ and $\exp$-factors in Eq.~\ref{Eq:average_nu} vs.~${\cal T}_0$. Figure~\ref{lobezeros_nu2nd}c shows the variation of the inverse node spacing $\e_L/[e(V_2-V_1)]$ extracted from Fig.~\ref{lobezeros_nu2nd}a. Instead of a step function (dotted line) expected from the elementary argument leading to Eq.~\ref{Eq:average_nu}, a rounding of the step is found that results from quantum fluctuations of $n$ and agrees well with our numerical calculation (full line). On the other hand, the Gaussian truncation of the FCS predicts a linear decay from 1 to 0 (dashed line) rather than a step. Again, the absence of higher visibility nodes below ${\cal T}_0=1/2$ demonstrates the essential contribution of the higher cumulants contained in Eq.~\ref{FCS}. Moreover, the abrupt decay of $\e_L/[e(V_2-V_1)]$ at the transition point ${\cal T}_0=1/2$ suggest to consider this quantity as a normalized order parameter for the non-equilibrium phase transition predicted in Ref.~\onlinecite{sukh2}.

In Fig.~\ref{lobezeros_nu2nd}d we display the peaked behavior of the dephasing rate $V^{-1}_D({\cal T}_0)=(e/2L)\cdot\ln|2{\cal T}_0-1|$ defined by Eq.~\ref{Eq:average_nu}.  The points were extracted from from Fig.~\ref{lobezeros_nu2nd}b for ${\cal T}_0>0.5$, and the exponential tails of $\nu(\Vdc)$ in Fig.~\ref{lobechange_new} for ${\cal T}_0\leq0.5$. We see that $V_D^{-1}({\cal T}_0)$ shows indeed a pronounced peak at the transition point ${\cal T}_0=0.5$ which decays in a way that is quantitatively reproduced by the theory, using again the same value for the velocity of dipolar plasmons determined from Fig.~\ref{char_lobes}c. The singular dephasing rate at the transition point reflects the dominance of the non-Gaussian noise at the non-equilibrium phase transition.

We have seen that our model, despite being simple, is able to explain quite
sophisticated phenomena such as the lobes in the visibility of AB oscillations
and the non-Gaussian noise induced phase transition. Therefore, it seems
that our model is able to grasp the essential physics of dephasing in MZIs.
However, it remains to discuss to what extent the predictive power of our
theory is robust against various modifications of the model. In this connection
we wish to mention the recent paper Ref.~\onlinecite{mirlin2}, which investigates AB oscillations
in the MZI at integer filling factors using a model with a different form of the
interaction potential. Namely, in contrast to our model, where the interaction
extends outside the interferometer and is approximated by a short-range
potential, the authors of Ref.~\onlinecite{mirlin2} consider an opposite extreme where
the the interaction is localized inside the MZI between two QPCs, and the
potential is approximated by a ``capacitance model'', i.e., it is maximally
long-range.

Concerning the lobe structure, the results of Ref.~\onlinecite{mirlin2} for {\it ff}$\,=2$ and for strong interaction
limit fully agree with our findings as well as with the results of the earlier paper Ref.~\onlinecite{sukh1},
where our present model has been thoroughly analyzed. We note, that
the authors also predict lobes for the case of {\it ff}$\,=1$. However, their moderate
interaction limit, where the lobes are predicted, is not quite consistent, because
the the long-range Coulomb interaction is effectively always strong. Moreover,
Ref.~\onlinecite{mirlin2} neither makes any attempt to interpret the physical origin of predicted lobes, nor it discusses to what extent they are robust against the variation of the transparencies of the QPCs of the MZI. In any case, the predicted behavior at {\it ff}$\,=1$ seem to have no relation to the phenomenon investigated in the present paper.

In conclusion, our work constitutes a detailed investigation of the effect of non-Gaussian noise on the visibility of interference in electronic interferometers. The lobe- and node structure of the visibility directly reflects the peculiar analytic structure of the generator of the full counting statistics of a quantum point contact. It provides first experimental evidence of a singularity in the FCS that is intrinsic to binomial statistics, and that induces a novel type of non-equilibrium phase transition. In addition, the excellent overall agreement of the observed behavior of the visibility with the predictions of the plasmonic edge model provides firm evidence that the latter captures the essential physics behind the complex interference phenomena observed in Mach-Zehnder interferometers with two co-propagating edge channels. \\[2mm]
\emph{Acknowledgements:} We want to thank K.~Kobayashi, H.~S.~Sim and S.~Ludwig for fruitful discussions and H.-P.~Tranitz for growing the GaAs/AlGaAs-material. The work was funded by the DFG within the SFB631 "Solid state quantum information processing", and by the Swiss NSF.\\[2mm]
\emph{Contributions:}
C.S. and E.V.S. conceived the experiment. L.V.L. and A.H. fabricated the samples, conducted the measurements, and analyzed the data. I.P.L. and E.V.S. provided the numerical data.  All authors contributed to the writing of the manuscript.



\begin{thebibliography}{35}
\bibitem{neutrino}
In our case,
 the flavor quantum number is replaced with Landau level number,
and the neutrino mass eigenstates are replaced with plasmon eigenmodes.
\bibitem{footnote_epsilon_L}
This definition of $\e_L$ agrees with that of Ref.~\onlinecite{Leonid2}, and differs from that of Ref.~\onlinecite{sukh2} by a factor of $2\pi$.
\end{thebibliography}
\end{document}


\title{Methods and Supplementary information for ``Counting Statistics and Dephasing Transition in an Electronic Mach-Zehnder Interferometer'}
\author{A. Helzel }
\author{L. V. Litvin }
\affiliation{\mbox{Institut f\"{u}r experimentelle und angewandte Physik, Universit\"{a}t  Regensburg, D-93040 Regensburg, Germany}}
\author{I. P. Levkivskyi}
\affiliation{\mbox{Theoretische Physik, ETH Zurich, CH-8093 Zurich, Switzerland}}
\affiliation{Bogolyubov Institute for Theoretical Physics, 03680 Kiev, Ukraine}
\author{E. V. Sukhorukov}
\affiliation{\mbox{D\'{e}partement de Physique Th\'{e}orique, Universit\'{e} de Gen\`{e}ve, CH-1211 Gen\`{e}ve 4, Switzerland}}
\author{W. Wegscheider }
\affiliation{\mbox{Laboratorium f\"{u}r Festk\"orperphysik, HPF E 7, ETH Z\"urich, 8093 Z\"urich, Switzerland}}
\author{C. Strunk}
\affiliation{\mbox{Institut f\"{u}r experimentelle und angewandte Physik, Universit\"{a}t  Regensburg, D-93040 Regensburg, Germany}}
\date{\today}
\pacs{73.23.Ad, 73.63.Nm}
%
%
\maketitle

\section{Experimental Methods}

%
\begin{figure}[tt]
\includegraphics[width=80mm]{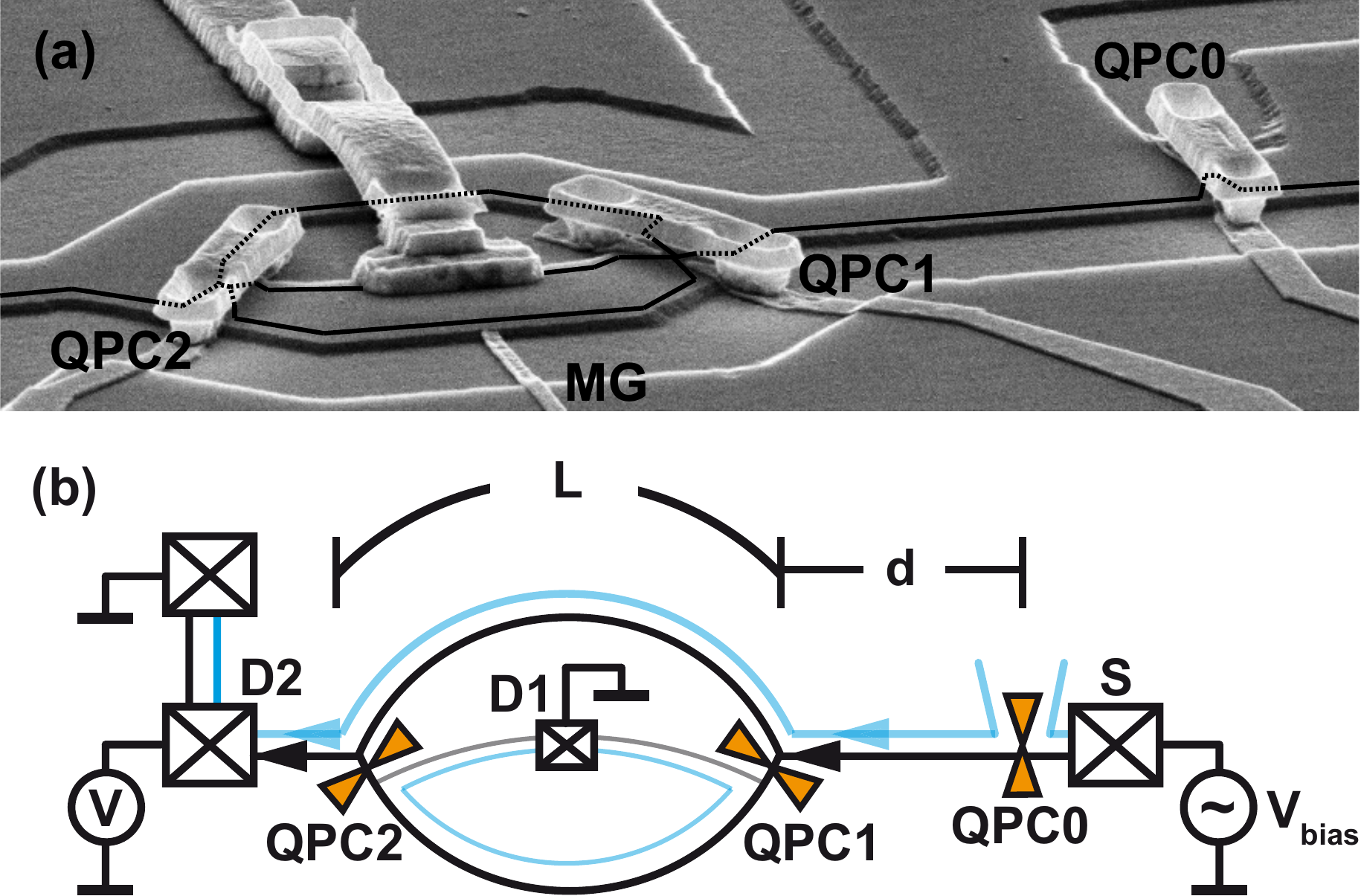}\\
%
\caption{\textbf{SEM picture and sketch of the sample:} (a) Scanning electron micrograph of the sample with marked gates QPC~0, QPC~1, QPC~2 and the modulation gate MG. The black line represents the interfering edge channel with QPCs~1 and 2 are set to half transmission of the outer edge channel. QPC~0 fully reflects the inner and partially transmits the outer edge channel. (b) Schematic of the relevant edge channels in the MZI in addition to the source \textbf{S}, the drains \textbf{D1} and \textbf{D2} and the QPCs~1, 2 and 0. The black line is the outer edge channel carrying interfering electrons and the light blue line represents the inner edge channel. QPC~0 is adjusted to reflect the inner edge channel (blue). Important sample dimensions are the length of the interferometer arms $L$ and the distance $d$ between QPC~0 and 1.}
\label{sample}
\end{figure}

The results are obtained on two samples made from different wafers. Sample A was structured in a modulation doped GaAs/Ga$_{x}$Al$_{1-x}$As heterostructure with a two dimensional electron gas (2DEG) 90\,nm below the surface. The 2DEG density and mobility are $n=2.0\times10^{15}$\,m$^{-2}$ and $\mu=206$\,m$^{2}/$(Vs) at 4\,K on the unpatterned wafer. Sample B was patterned in another heterostructure also with a depth of the 2DEG of 90\,nm, but with $n=2.1\times10^{15}$\,m$^{-2}$ and $\mu=289\,$m$^{2}/$(Vs) at 4\,K on the unpatterned wafer. The exact patterning procedure is described in Ref.~\onlinecite{Leonid1}. The arm's length $L$ of the interferometers is estimated to be 6.5\,$\mu$m for sample A and 8.7\,$\mu$m for sample B. The structures contain not only the MZI, i.e. QPC~1 and 2 and two drains, but also an additional quantum point contact (QPC0) between source and MZI. This QPC0 is situated a distance $D=5\,\mu$m for sample A and $8\,\mu$m for sample B in front of QPC~1 (see Fig.~\ref{sample}).

\begin{figure}[tt]
\includegraphics[width=75mm]{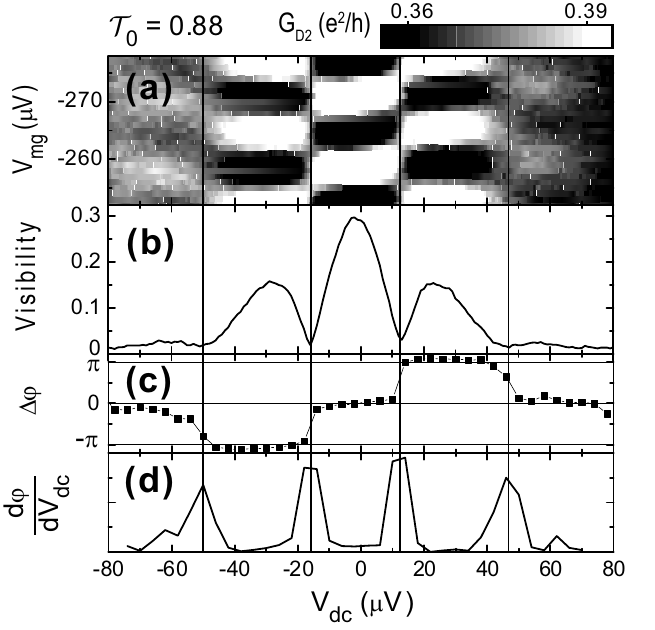}
\caption{\textbf{Raw and processed data:} (a) Gray scale plot of the measured conductance $G(V_{\textrm{dc}},V_{\textrm{mg}})$ for $\mathcal{T}_0=0.88$. Checker-board pattern of dark and bright regions reveals central lobe and two additional lobes on each side. (b) The visibility of the in (a) measured oscillatory conductance extracted according to Eq.~\ref{vis}. (c) The AB-phase shift $\Delta\varphi$ with respect to $V_{\rm dc}=0$ obtained from fits of traces $G(V_{\textrm{mg}})$ with Eq.~\ref{sine}. Constant phase inside lobes and jumps of $\pi$ at nodes can be seen. (d) The numerical derivative of $\varphi$ further highlights the position of the nodes as pronounced peaks.}
\label{example}
\end{figure}

A standard lock-in technique ($f\sim300$\,Hz) was used to measure the output current via voltage drop between terminal D2 and another (grounded) ohmic contact. An ac voltage of 1\,$\mu$V plus a dc voltage $V_{\textrm{dc}}$  were applied to measure the differential conductance  $G(V_{\textrm{dc}})=dI(V_{\textrm{dc}})/dV_{\textrm{dc}}$. By comparing temperature dependence of the visibility in Ref.~\onlinecite{Leonid2} (the bath temperature of the dilution refrigerator was measured with a RuO$_x$ thermometer) with that in Ref.~\onlinecite{hashisaka}, where the electron temperature in the interferometers was measured directly via the thermal noise, we estimate the electron temperature in the present experiment to be close to 30\,mK. Sample A was measured at a magnetic field of 4.73\,T (${\it ff}=1.7$) and with ideal configuration of the QPCs a (maximum) visibility of 65\,\% at $V_{\textrm{dc}}=0$ is achieved, sample B at 4.5\,T (${\it ff}=1.8$) with a maximum visibility of 33.5\,\%.

\subsection{Analysis:}
We measured $dI/dV_{\textrm{dc}}=G$ vs.~$V_{\textrm{dc}}$ for a range of modulation gate voltage $V_{\textrm{mg}}$. An example of raw data can be seen in Fig.~\ref{example}a. The sinusoidal oscillations for ramping $V_{\textrm{mg}}$ and the decaying oscillations for $V_{\textrm{dc}}$ are well-defined. In the raw data the multiple side lobes can be seen as a chess board pattern that fades out at larger $V_{\textrm{dc}}$. We extract the differential visibility
\begin{equation}\label{vis}
\nu(V_{\textrm{dc}})=\frac{G_{max}(V_{\textrm{dc}})-G_{min}(V_{\textrm{dc}})}{G_{max}(V_{\textrm{dc}})+G_{min}(V_{\textrm{dc}})}
\end{equation}
of the interference pattern at each bias voltage $V_{\textrm{dc}}$ (Fig.~\ref{example}b). The Aharonov Bohm phase and the visibility in Fig.~2 of the main part are determined from sine fits of modulation gate traces at certain $V_{\textrm{dc}}$ relative to the trace at zero bias (Fig.~\ref{example}c).

\begin{figure}[t]
\includegraphics[width=85mm]{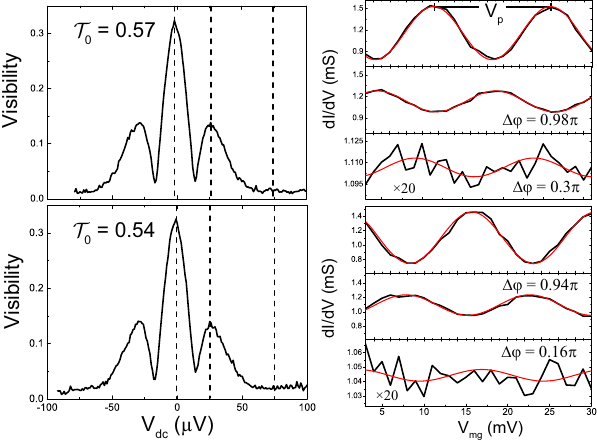}
\caption{\textbf{Sine fits of gate traces:} On the left the visibilities vs. $V_{\textrm{dc}}$ for $T_{\textrm{QPC0}}=0.57$ (top) and $0.54$ (bottom) are displayed. On the right are according modulation gate traces (black line) of the marked $V_{\textrm{dc}}$ (dashed vertical lines in the left plots) at the center of the lobes and their sine fits (red line). For $V_{\textrm{dc}}\approx 75\,\mu V$ only residual oscillations buried in noise are present.}
\label{osci}
\end{figure}

We fit the measured modulation gate traces to
\begin{equation}
G(V_{\rm mg})\ =\ G_{av}+G_\mathrm{osc}\sin(V_{\rm mg}/V_p+\Delta\varphi)
\label{sine}
\end{equation}
with the four parameters $G_{av}$, $G_\mathrm{osc}$, $V_p$, and $\Delta\varphi=\varphi(0)-\varphi(V_{\rm dc})$. For a measurement of each transmission $\mathcal{T}_0$ we fit the period $V_p$ only once for the large oscillations at zero bias and then keep it fixed for the other traces of one measurement. Examples of typical modulation gate traces and the corresponding fit curves can be seen in Fig.~\ref{osci} for transmissions 0.57 and 0.54. In this way we trace the phase evolution at the frequency of interest even if it is nearly buried in noise. This becomes crucial for very small oscillation amplitudes. From such fits we determined $\nu_{2nd}$ in Fig.~4b of the main part for transmissions $\mathcal{T}_0$ close to 0.5.

Fig.~\ref{example}d shows the numerical derivative of $\varphi(V_{\rm dc})$ to highlight the phase jumps at the nodes, as it is used in Fig.~3, main part. For the evaluation of the node positions $V_m$ and the height of the second side lobe in Fig.~4 of the main part, both visibility and phase evolution are analyzed to determine the node positions.

\begin{figure}[t]
\includegraphics[width=85mm]{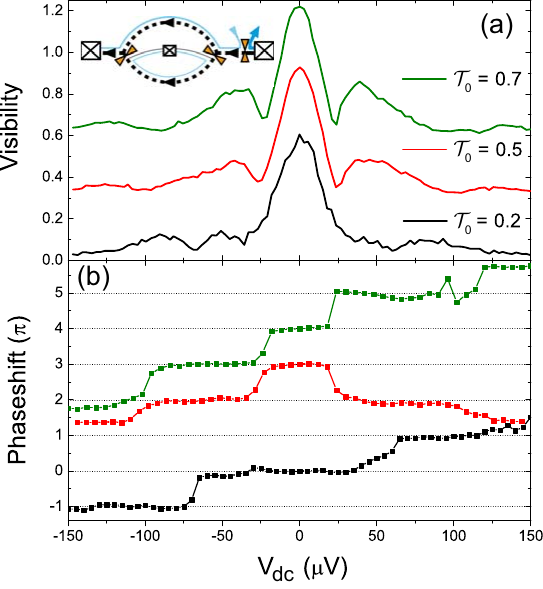}
\caption{{\bf Evolution of the lobe structure in sample A:} Visibility (a) and phase shift (b) vs.~dc bias voltage for various transmissions of QPC0 for sample A. For $\mathcal{T}_0\geq0.5$ multiple side lobes are present in the visibility and the phase evolution, similar to sample B. For $\mathcal{T}_0=0.2$ we see a wide central lobe and only single side lobes. The position of the nodes is clearly visible in the phase behavior by jumps and we can deduce the lobe structure.}
\label{sampleA}
\end{figure}

\section{Characterization of the interferometers:} Two samples are studied, which differ in interferometer arm length $L$ and distance $d$ (see Methods). The maximal two-terminal conductance is $\approx 2e^2/h$, corresponding to two transmitted edge channels. We use QPC0 to selectively bias the outer edge channel, while the source terminal of the inner channel is left grounded (${\mathcal T}_0=1$). QPCs 1 and 2 are set to reflect the inner channel, implying that the interference takes place only in the outer edge channel. At zero bias we reach maximal interference visibilities ($\nu_0$) of 65\,\% in sample A and 33.5\,\% in sample B. Besides the maximum visibility $\nu_0$ the lobe periodicity $V_0$ is different for the samples. Both parameters depend on the magnetic field, the temperature and the arm length $L$ of the interferometer.\cite{roche2,Leonid2,sukh2}

The  visibility $\nu_0$ at zero dc bias decreases exponentially with $L$, $\nu \propto \exp(-2L/l_{\varphi})$, with the coherence length $l_{\varphi}\propto T^{-1}$, similar to the data in Refs.~\onlinecite{roche2} and~\onlinecite{Leonid2}. The different maximum visibilities of the two samples result from the different sizes $L$, affecting both $\nu_0$ and $V_0$. The normalized visibility pattern $\nu(V_\mathrm{dc}/V_0)/\nu_0$ in Fig.~1c of the main part turns out to be independent of $L$. We can conclude that for $\mathcal{T}_0=1$ the differences between both samples are controlled by only one parameter, i.e., the $L$-dependent characteristic energy $\varepsilon_L=2\pi\hbar v/L$. The presence of visibility nodes is even more clearly visible in the evolution of the Aharonov-Bohm phase. In the visibility only two side lobes can be seen clearly, whereas the analysis of the residual oscillations at high bias voltages by sinusoidal fitting (see methods) display one more phase jump, revealing a third side lobe in sample B.
\begin{figure}[t]
\includegraphics[width=85mm]{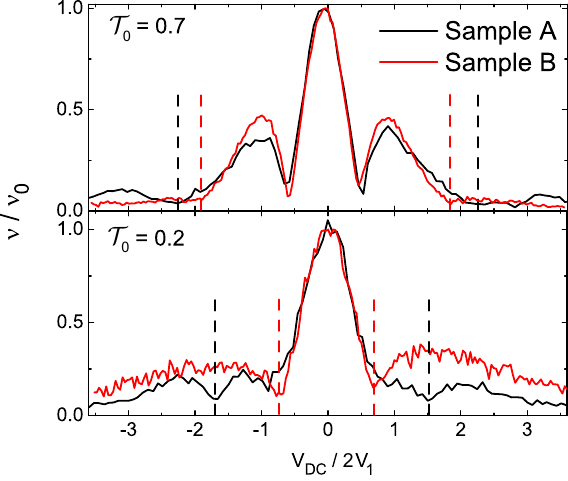}
\caption{\textbf{Differences between samples A and B:} The visibility for $\mathcal{T}_0=0.7$ and $\mathcal{T}_0=0.2$ of both samples are shown. The dashed lines mark the position of the second nodes (upper panel) and first nodes (lower panel) for sample A (black) and B (red). Though for $\mathcal{T}_0=1$ the curves for sample A and B coincide after scaling by $\nu_0$ and $V_0$, this is different for lower transmission of QPC0. }
\label{scaleAB}
\end{figure}

\section{Comparison of samples A and B}
In the main part we mainly display data obtained on sample B. Here we  present an overview of the lobe structure for sample A in Fig.~\ref{sampleA}. The qualitative behavior in sample A is similar to sample B, i.e., the positions $V_m$ of the multiple side nodes (see Fig.~\ref{sampleA}a) and the phase jumps (see Fig.~\ref{sampleA}b) remain fixed for $\mathcal{T}_0>0.5$ and the location $V_1$ of the single side node increases for $\mathcal{T}_0<0.5$. Such behavior is also seen in the numerical derivative of $\varphi(V_{\rm dc})$ (Fig.~3 of the main part).
  For $\mathcal{T}_0=1$ we can match the visibility curves for the two samples very well by normalization with respect to the node spacing $V_0$ and the zero-bias visibility $\nu_0$ (see Fig.~1c in the main part).

In contrast, for lower transmissions $\mathcal{T}_0<1$ of QPC0 differences between sample A and B remain, which originate from the operation of QPC0 as a source of current noise at the interferometer input. In Fig.~\ref{scaleAB} the discrepancies are shown for two exemplary transmissions $\mathcal{T}_0=0.7$ and $\mathcal{T}_0=0.2$. The voltages are scaled with respect to $2V_1$ extracted from the trace with $\mathcal{T}_0=1$. The measurement of the phase shift in Fig.~\ref{sampleA}b allows an unambiguous determination of the node positions $V_m$, which are marked by the dashed lines in Fig.~\ref{scaleAB}. Because of the scaling the first visibility nodes match, but the second visibility nodes disagree for $\mathcal{T}_0=0.7$. Moreover, even the first visibility nodes of the two samples do not collapse for $\mathcal{T}_0=0.2$. In addition, a foot develops in $\nu$ near $V_{\rm dc}\lesssim V_1$.

\begin{figure}[t]
\includegraphics[width=75mm]{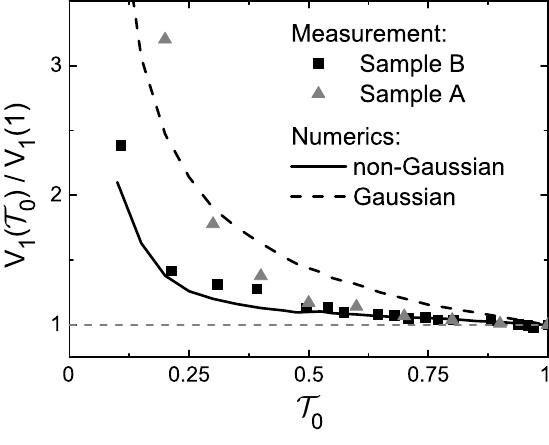}
\caption{{\bf Evolution of $V_1(\mathcal{T}_0)$:} The position of the first node $V_1$ follows the non-Gaussian prediction for large $\mathcal{T}_0$. For transmissions $\mathcal{T}_0<0.5$ the voltage $V_1$ of the first visibility node (dots) grows more rapidly for sample A  than for sample B, indicating a cross-over from non-Gaussian (solid line) to  Gaussian (dashed line) behavior, in particular for sample A.}
\label{V1}
\end{figure}

The overall variation of $V_1$ vs.~$\mathcal{T}_0$ for both samples is shown in Fig.~\ref{V1} together with theory curves for the Gaussian and non-Gaussian case. For lower transmissions the shift of $V_1$ with transmission is much stronger for sample A, when compared with sample B.

Below $\mathcal{T}_0=0.5$ a cross-over from non-Gaussian to Gaussian behavior is observed for sample A. This observation can be made plausible by the following argument: at finite voltages the statistics of the particle numbers transmitted through QPC0 is expected to be non-Gaussian.\cite{levitov}
It was shown in Ref.~\onlinecite{sukh4}, however, that a weak non-linearity in the spectrum of the plasmon modes $k(\omega)=\omega/v+\gamma \omega^2 \textrm{sign}(\omega)$ can suppress the contributions of higher order cumulants in the FCS generator for distances $L_g=1/(\gamma TV_{\rm dc})^2$ that strongly depend on the bias voltage $V_{\rm dc}$. Such a non-linear plasmon dispersion relation can also lead to a non-linear conductance of the QPC.
A  cross-over to Gaussian noise can result from a decrease of $L_g(V_{\rm dc})$ with larger $V_{\rm dc}$, or from sufficiently strong nonlinearity of the plasmon dispersion relation in sample A, which ensure $L_g<d$ already at small voltages. We checked carefully that sample B shows negligible nonlinearities in the current-voltage-characteristic of the QPCs. This is consistent with the observed non-Gaussian behavior of the visibility. On the other hand, we found that sample A has strong nonlinearities in the conductance. Because the arm length $L$ of interferometer A is smaller, the important energy $\varepsilon_L$ and thus the required voltages $V_m$ are larger in sample A (the ratio $L/d$ is similar in both samples). Together with the strong voltage dependence of $L_g$ this may explain the observed cross-over from non-Gaussian to Gaussian behavior of $V_1$ for $\mathcal{T}_0<0.5$ and $V_2$ for $\mathcal{T}_0>0.5$ in Figs.~\ref{scaleAB} and~\ref{V1}. The larger characteristic energy $\varepsilon_L$ of sample A makes it more prone to a suppression of higher order cumulants at larger voltages, when compared with sample B.

\begin{figure}
\includegraphics[width=85mm]{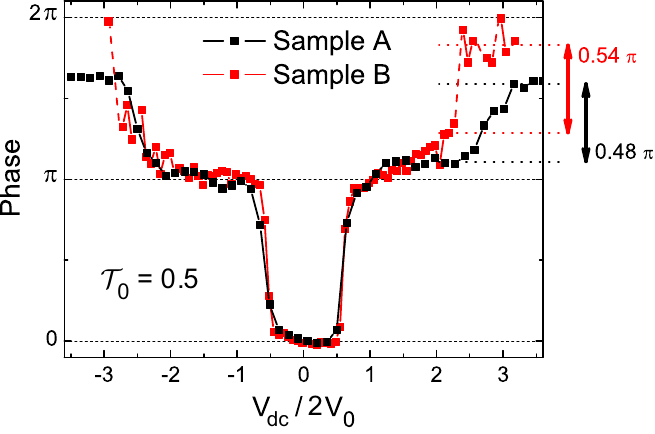}
\caption{\textbf{AB-phase evolution at the phase transition:} For $\mathcal{T}_0=0.5$ (measured for $V_\mathrm{dc}=0$) multiple side lobes are observed for sample A. In sample B only single side lobes are visible with a width similar to $\mathcal{T}_0>0.5$. The phase evolution shows in both samples a jump of $\pi$ at the first node. In sample A we clearly see a further increase of the phase that saturates near $3\pi/2$. The second side lobes are pronounced. Sample B shows a similar unexpected behavior.}
\label{phase05}
\end{figure}

From Ref.~\onlinecite{sukh2} it is expected that there are no multiple side lobes at $\mathcal{T}_0=0.5$. In contrast to this prediction we observe traces of a second side lobe in our experimental data at $\mathcal{T}_0=0.5$, in particular for sample A (see Fig.~\ref{sampleA} and Fig.~2, main part). Fig.~\ref{phase05} shows the measured phase shifts for both samples at this point. After the jump of the phase from $0$ to $\pi$ at the visibility node a further gradual shift of the phase towards higher values can be clearly seen.  The additional phase shift saturates at  $\pi/2$ for sample A. In sample B only residual traces of an additional side lobe are visible, which also do not obey phase rigidity. The fact that multiple side lobes are observed near $\mathcal{T}_0=0.5$ may be explained by a variation of ${\cal T}_0$ with $V_{\rm dc}$ that are consistent with the observed non-linearities of the QPCs of sample A,\cite{leonid3} and result in a transmission ${\cal T}_0$ at finite $V_\mathrm{dc}$, which slightly increases with $V_\mathrm{dc}$. On the other hand, the deviations from phase rigidity observed in Fig.~\ref{phase05} cannot be understood in this way.

\section{Theoretical method}
In the main part of the paper, we present an explanation of the lobe-type pattern
of visibility in electronic MZI in simple physical terms.
A more rigorous description of QH interferometers at filling factor 2, proposed in Ref.~\onlinecite{sukh1}, is
based on the so-called bosonization approach. Here, we summarize this approach
in order to support the elementary derivation given in the main part of the paper. For simplicity, we set $e=\hbar=1$
in the beginning and restore physical units at the end of calculations.

Our experiment addresses the physics of MZIs at low-energies compared to the Fermi energy and
at long distances compared to the magnetic length.
The low-energy spectrum of excitations of chiral edge states consists of the collective charge density
oscillation (plasmons). Using second quantization language, these excitations are described by
creation and annihilation operators, satisfying the bosonic commutation relations
\begin{equation}
[a^\dag_{s\alpha k},a_{s'\beta k'}] = \delta_{ss'}\delta_{\alpha\beta}\delta_{kk'},
\label{eq1}
\end{equation}
where $s = 1, 2$ enumerates two arms of
the interferometer, $\alpha = 1,2$ enumerates two Landau levels at filling factor 2,
and $k$ is the wave vector. Namely,
1D charge densities may be expressed as
\begin{equation}
\rho_{s\alpha}(x) = (1/2\pi)\partial_x\phi_{s\alpha}(x)
\label{eq2}
\end{equation}
in terms of boson fields
\begin{eqnarray}
\phi_{s\alpha}(x) &=&
\varphi_{s\alpha}+2\pi N_{s\alpha}x/W\nonumber\\
&+&  \sum_{k>0}\sqrt{2\pi/kW}(e^{ikx}a_{s\alpha k} + e^{-ikx}a^\dag_{s\alpha k}),
\label{eq3}
\end{eqnarray}
which satisfy the canonical commutation relations
$[\partial\phi_{s\alpha}(x),\phi_{s'\beta}(y)] = 2\pi i \delta_{ss'}\delta_{\alpha\beta}\delta(x-y)$.
Here, $W$ is the total size of the system (to be taken to infinity in the end of calculations),
and the operators $\varphi_{s\alpha}$ and $N_{s\alpha}$ are the so-called
zero modes, i.e, the modes with $k=0$: $N_{s\alpha} = \int dx\rho_{s\alpha}(x)$
is the total number of electrons in the channel $(s, \alpha)$, and operators $\exp[- i\varphi_{s\alpha}]$
increases this number by $1$.

The key idea of the bosonization technique is to express the electron creation operators
in terms of the boson fields:
\begin{equation}
\psi^\dag_{s\alpha}(x) = \exp[-i\phi_{s\alpha}(x)].
\label{eq4}
\end{equation}
With the help of Eqs.\ (\ref{eq1}-\ref{eq3}) one can check that (i)
so-defined operators obey {\em fermionic} anti-commutation relations; (ii) they create
local excitations with unit charge, i.e., they commute with charge density operators as
$[\rho_{s\alpha}(x), \psi^\dag_{s\alpha}(y)] = \delta(x-y)\psi^\dag_{s\alpha}(x)$.
When acting on the state of the interferometer, such operator
first increases the total number of electrons $N_{s\alpha}$ by one, and second, it creates a bunch of
plasmon excitations localized near the point $x$, as can be easily seen from Eq.\ (\ref{eq3}).
These two effects are reflected in Eq.~2 in the main text.

The convenience of the bosonization approach is in the fact that the Hamiltonian of interacting 1D electrons
is quadratic in terms of the boson fields and can be easily diagonalized.
In particular, it has been shown in Ref.~\onlinecite{sukh1} that the Hamiltonian of electrons
interacting via the short-range potential $U(x,y) = U\delta(x-y)$ can be written as:
\begin{equation}
\mathcal{H} = \frac{1}{2}\sum_{s,\alpha,\beta}
\int_0^W dx V_{\alpha\beta}\rho_{s\alpha}(x)\rho_{s\beta}(x),
\label{eq5}
\end{equation}
where the inverse ``capacitance'' matrix $V_{\alpha\beta} = U + 2\pi v_F\delta_{\alpha\beta}$ contains
the Fermi sea contribution with the Fermi velocity $v_F$. It is easy to see that this Hamiltonian
can be rewritten in the diagonal form:
\begin{equation}
\mathcal{H} = \int_0^W \frac{dx}{4\pi} \left[u (\partial_x\tilde{\phi}_{s1}(x))^2 + v (\partial_x\tilde{\phi}_{s2}(x))^2\right]
\label{h-sup}
\end{equation}
in terms of the charge and dipole modes
$\tilde{\phi}_{s1,2}(x) = [{\phi_{s1}(x)\pm\phi_{s2}(x)}]/{\sqrt{2}}$.
Note that the velocity of the charge mode $u = U/\pi + v_F$ is much larger then
the velocity of dipole mode $v = v_F$ in the limit of the strong interaction $U\gg v_F$,
which applies, e.g., for Coulomb interactions screened at relatively long distances.
The new plasmon operators
\begin{equation}
\tilde{a}_{s1k} = \frac{1}{\sqrt{2}}(a_{s1k}+ a_{s2k}),\;
\tilde{a}_{s2k} = \frac{1}{\sqrt{2}}(a_{s1k}- a_{s2k})
\label{eq7}
\end{equation}
have a simple physical meaning: They create and annihilate charge and dipole plasmon
excitations with the wave vector $k$.

Next, we note that the zero mode contribution
$\sum V_{\alpha\beta}N_{s\alpha}N_{s\beta}/2W$ to the Hamiltonian (\ref{eq5})
can be interpreted as an energy of a capacitor with capacitance matrix $W\cdot V^{-1}_{\alpha\beta}$.
Using this fact one can find the average value of the charge operators in terms of electro-chemical potentials
$\Delta\mu_\alpha$. In particular, the total number of electrons in the upper outer channel reads:
$N_{U1} = W\sum \cdot V^{-1}_{1\alpha}\Delta\mu_\alpha \simeq W\Delta\mu_1/4\pi v$, for $\Delta\mu_2=0$, and
where we have neglected small contribution $\sim 1/u$.
Thus, according to Eqs.~(\ref{eq3}) and (\ref{eq4}),  the excitations created by electron tunneling
acquire the following phase from zero modes (restoring physical units and setting $\Delta\mu_1=eV_\mathrm{dc}$)
\begin{equation}
\delta\varphi_{0} = -W\Delta\mu_1/4\pi \hbar v \cdot 2\pi L/W = -eV_\mathrm{dc} L/2\hbar v,
\label{eq10'}
\end{equation}
which explains Eq.~5 in the main text.

Let us now consider the dynamical phase acquired by the plasmons. From the Hamiltonian
(\ref{h-sup}) it follows that:
\begin{equation}
\tilde{a}_{s1k}(t) = e^{-iukt}\tilde{a}_{s1k}, \hspace{12pt} \tilde{a}_{s2k}(t) = e^{-ivkt}\tilde{a}_{s2k}.
\label{eq8}
\end{equation}
These relations determine the time evolution of the electron operator (\ref{eq4}). On the other hand, the
wave function overlap introduced in the main part of the paper my be written as
\begin{equation}
\langle\psi_{N+1}|\psi'_{N+1}\rangle\propto\int dt e^{\Delta\mu_1t}\langle N|\psi_{U1}(0,0)\psi^{\dagger}_{U1}(L,t)|N \rangle,
\label{eq9}
\end{equation}
where the time integral projects an electron onto the energy $\Delta\mu_1$, with which it is injected.
The similar contribution from the lower arm of the interferometer has been omitted for the sake of simplicity of
the argument.

The complication in the next step arises because each electron operator on the right hand side of (\ref{eq9}) generates
an infinite number of terms, when expanded in the plasmon operators, which can be schematically expressed as following:
$\langle\psi_{N+1}|\psi'_{N+1}\rangle\propto\sum_{\{k,k'\}} C_{\{k\}}C_{\{k'\}}e^{-i(K+K')L}\delta(\Delta\mu_1+Ku+K'v)$, where $C_{\{k\}}$
and $C_{\{k'\}}$ are the plasmon correlation functions for the sets of wave numbers $k_i$ and $k'_i$, and $K=\sum_i k_i$, $K'=\sum_i k'_i$.
Here we have used
Eqs.\ (\ref{eq3}), (\ref{eq4}), (\ref{eq8}), and integrated over time $t$ to obtain the energy conserving delta-function.
It is easy to see that in the large-$L$ limit the summation over all possible plasmon excitations leads to fast oscillations
in the above expression and to the suppression of corresponding contributions. However, two terms in this sum, the separate
contributions of the dipole and charge mode, constitute an exception:
$\langle\psi_{N+1}|\psi'_{N+1}\rangle\propto\sum_{\{k\}} C_{\{k\}}e^{-iKL}\delta(\Delta\mu_1+Ku)+
\sum_{\{k'\}} C_{\{k'\}}e^{-iK'L}\delta(\Delta\mu_1+K'v)$. Thus, as a consequence of the linear spectrum of plasmons
(and of the chirality of the system), we immediately arrive at the expression
$\langle\psi_{N+1}|\psi'_{N+1}\rangle\propto e^{i\Delta\mu_1L/u}+e^{i\Delta\mu_1L/v}$, which justifies our simplified
approach in the main part of the paper, leading to Eq.~(4).
In the limit $u\gg v$, restoring physical units and setting $\Delta\mu_1=eV_\mathrm{dc}$, the relative phase shift due to the dipole mode reads
\begin{equation}
\delta\varphi_{\rm d} = eV_\mathrm{dc} L/\hbar v.
\label{eq9'}
\end{equation}
Note, that the universality of the ratio
$\delta\varphi_{\rm d}/\delta\varphi_{0}=-2$, which follows from strong interactions
at QH edge, explains the origin of the phase rigidity and phase transition phenomena
observed in our experiment.

\begin{figure}[t]
\includegraphics[width=85mm]{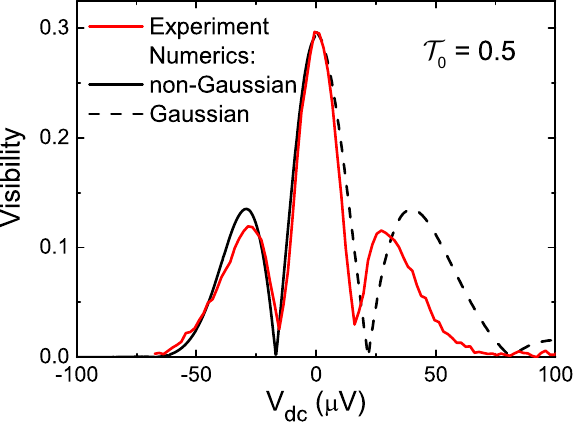}\\
\vspace{4mm}
\includegraphics[width=85mm]{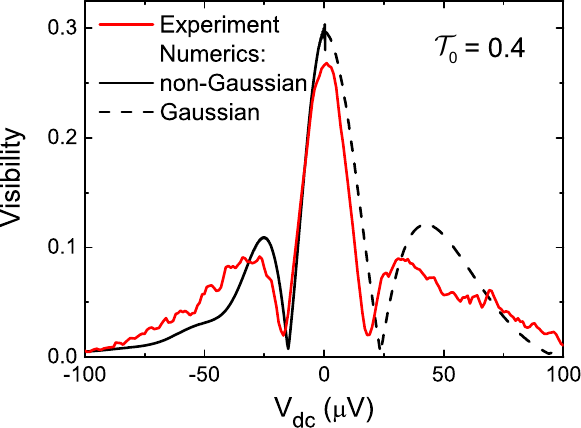}
\caption{\textbf{Comparison to theory} for $\mathcal{T}_0=0.5$, and $\mathcal{T}_0=0.4$ of sample B. In the fits of Gaussian (right half of graphs) and non-Gaussian (left half of graphs) noise the only free parameters are $\nu_0$ and $V_2$ at $\mathcal{T}_0=1$. The better agreement to the non-Gaussian prediction is obvious.}
\label{gauss05}
\end{figure}

\section{Numerical results for Gaussian and non-Gaussian noise versus experimental data}
In the main part of this work we illustrated the qualitative signatures of the phase transition when going from $\mathcal{T}_0>0.5$ to $\mathcal{T}_0<0.5$. Here we want to go more into detail of the expected differences in the visibility characteristics predicted for Gaussian and non-Gaussian noise. In Fig.~\ref{gauss05} we compare two traces of $\nu(V_{\textrm{dc}})$ for $\mathcal{T}_0=0.5$ and $\mathcal{T}_0=0.4$ with the results of the numerical calculations described in the following.

Ref.~\onlinecite{sukh2} provides analytically-derived large-bias asymptotics of the visibility of AB oscillations in Gaussian and non-Gaussian regimes. These asymptotics are based on the Levitov-Lesovik formula \cite{levitov} for the long-time behavior of the FCS generator of tunneling currents at the QPC0.
However, large-bias asymptotics are not sufficient for a direct
comparison with our experimental results.
For this reason, we follow the approach proposed in Ref.~\onlinecite{sukh2} and
evaluate the FCS generator numerically.
To this end we use the determinant representation of the FCS generator
(see Refs.~\onlinecite{levitov,bogoliubov}) which allows one to express this generator
as a determinant of a single-particle operator:
\begin{multline}\label{eq:integral}
\langle e^{i\lambda N(t)}e^{-i\lambda N(0)}\rangle\ =\ \\
\det[1-f(\varepsilon)+\exp(i\lambda U(t)\otimes S(\varepsilon))f(\varepsilon)],
\end{multline}
where $f(\varepsilon)$ is the energy distribution function, $U(t)$ is the projector on the time interval $[0,t]$, and $S(\varepsilon)$ is the scattering matrix of the QPC0.

We introduce a finite bandwidth for the electrons in the incoming channels
of QPC0 and fix it to be 4000 times larger than the level spacing.
Thus, we reduce the problem of finding the FCS generator to the
evaluation of the determinant of a finite matrix of the size 4000 by 4000.
The evaluation of such a determinant as a function of time $t$ and of the
transparency $\mathcal{T}_0$ can be trivially parallelized and has been done using
the Blue Gene/P machine \cite{bg}. Then,
we evaluate numerically the integral in the Eq.~3 of Ref.~\onlinecite{sukh2},
which connects the visibility and the FCS generator via Eq.~7
of Ref.~\onlinecite{sukh2}, and find the visibility as a function of the voltage
bias and of the transparency.

To compare this numerical data to the experiment we determine $\nu_0$, from the data of $\mathcal{T}_0=1$, and the position of the $2^{nd}$ node $V_2$ at $\mathcal{T}_0=1$ as in Fig.~4 in the main part.
Fig.~\ref{gauss05} shows the bias dependent visibility for $\mathcal{T}_0=0.5$ and $0.4$ in sample B with the numerical calculation for Gaussian and non-Gaussian noise. As one can see the nodes for the Gaussian prediction are expected for larger voltages as in the measurement and an additional side lobe should be present with a height that should be measureable. The curve for the non-gaussian prediction fits much better and the only small discrepancy is the height of the side lobe. This observation suggest again a strong, almost diverging, dephasing  characteristic for the non-Gaussian noise distribution expected after QPC0. The situation is similar for $\mathcal{T}_0=0.4$ -- multiple side lobes and position of nodes of the Gaussian prediction do not fit the measurement.

In conclusion, the two parameters $\nu_0$ and $V_0$ determined independently at $\mathcal{T}_0=1$ fix the whole set of visibility curves calculated for Gaussian and non-Gaussian noise at different $\mathcal{T}_0$. The experimental data agree much better with the non-Gaussian than with the Gaussian curves for all transmissions $\mathcal{T}_0$. This provides  striking evidence for the noise-induced phase transition proposed in Ref.~\onlinecite{sukh2}.

\section{Quantum measurements, entanglement, dephasing, and non-Gaussian noise}

It is instructive to consider the relation of the phase transition phenomena observed in our experiment
to the quantum measurement problem. 
In this section, we show that the comparison of the MZ interferometer to a quantum two-level system, 
although not
being precise, leads nevertheless to an important conclusion that the origin of the observed phenomenon lies 
in the perfect entanglement between electrons injected through the QPC0 toward the interferometer and
those in the superposition of occupying the upper or the lower arm of the interferometer.

\begin{figure}[t]
\includegraphics[width=85mm]{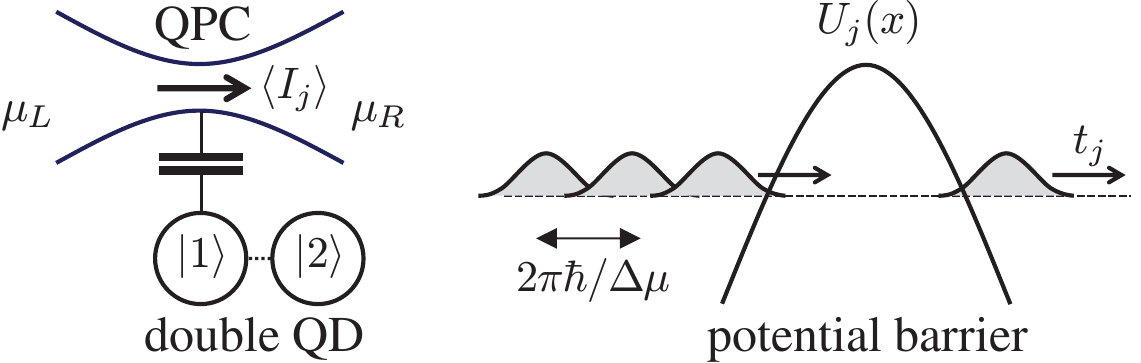}\\
\caption{A double quantum dot (QD) capacitively coupled to a QPC as an elementary example of a quantum measurement 
setup is schematically shown on the left. The QPC connects two electron reservoirs, biased with the electro-chemical
potential difference $\Delta\mu=\mu_R-\mu_L$. This causes the charge current through the QPC, which takes two
average values $\langle I_j\rangle$, $j=1,2$, depending on whether the state $|1\rangle$ or $|2\rangle$ of the double
QD is occupied. The simplified microscopic picture of the electron transport through the QPC is shown on the right. 
Electrons at Fermi level, incident 
from the left reservoir, collide with the potential barrier of the QPC, $U_j(x)$. Because of the capacitive coupling to the double QD, the potential $U_j(x)$ depends on the state $|j\rangle$ of the double QD. 
Therefore, the transmission and reflection coefficients, $t_j$ and $r_j$,
and as a consequence, the average current through the QPC $\langle I_j\rangle$, depend on the state of the double QD.}
\label{qms}
\end{figure}

A two-level system, e.g., a double quantum dot (QD) or a spin of electron, which interact with electrons in a QPC 
(see Fig.\ \ref{qms})
is an archetypal example of a quantum measurement setup.
This situation is relatively easy to describe theoretically,
and nevertheless, it contains essential physics. It also illustrates a dual character of a quantum measurement process.
On one hand, one may consider the QPC as a detector of the state of the two-level system. Namely, by applying the electro-chemical potential difference $\Delta\mu$ to the QPC, one generates the current, which on the time scale $2\pi\hbar/\Delta\mu$ or longer, depending on the character of coupling between two systems,
acquires one of the two values $I_1$ or $I_2$, corresponding to the final states of the two-level system.
At the same time, in the course of the measurement process, the current noise of the QPC dephases the 
quantum state of the two-level system, i.e., suppresses the off-diagonal elements of its density matrix. 
This leads to the idea\cite{levitov} to operate the setup in a dual mode, where the two-level system serves as 
a detector of the noise of the QPC: The off-diagonal elements of the density matrix of the two-level system
can be used as a measure of the FCS of the QPC's current noise.

The advantage of this gedanken experiment is that it can be theoretically described  using the single-particle scattering
approach. Here, we follow the analysis of Ref.\ \onlinecite{Averin}, 
allowing us to account for strong coupling between the QPC and the 
two-level system. Let us assume that the two-level system is initially in the pure quantum 
state $\sum_jc_j|j\rangle$, $j=1,2$. The interaction between the QPC and the two-level system
affects the QPC's potential barrier $U_j(x)$, which is experienced by the electrons incident 
at the QPC from the left reservoir (see Fig.\ \ref{qms}).
The potential barrier reflects an incident electron $|{\rm in}\rangle$ to the left outgoing state $|L\rangle$
with the amplitude $r_j$ and transmits it to the right outgoing state $|R\rangle$ with the amplitude $t_j$.
 As a result, the initial uncorrelated  state of the whole system $ |{\rm in}\rangle
\otimes\sum_jc_j|j\rangle$ evolves to the final pure state: 
\begin{equation}
|\psi\rangle=\sum_jc_j
\left(r_j|L\rangle+t_j|R\rangle\right)
\otimes|j\rangle.
\label{state}
\end{equation}
Taking a partial trace over electronic states 
$|L\rangle$ and $|R\rangle$, one finds that the initial reduced density matrix of the two-level system, $\rho_{jk}(0)=c_jc_k^*$,
evolves to the following final state
\begin{equation}
\rho_{jk}(1)=\rho_{jk}(0)\cdot (t_jt_k^*+r_jr_k^*)
\label{onepass}
\end{equation}
after the passage of one electron through the QPC. Applying this step $N$ times, one finds the reduced density matrix
after the passage of $N$ electrons:
\begin{equation}
\rho_{jk}(N)=\rho_{jk}(0)(t_jt_k^*+r_jr_k^*)^N.
\label{Npass}
\end{equation}
Thus, if electrons arrive with the rate $\Omega=\Delta\mu/2\pi\hbar$, the off-diagonal elements evolve as
\begin{equation}
\rho_{12}(t)=\rho_{12}(0)e^{h t},\quad \mbox{where}\; h=\Omega\ln(t_1t_2^*+r_1r_2^*).
\label{evolve} 
\end{equation}
It takes time of the order of $1/\Omega$ to resolve two average current levels $\langle I_j\rangle=\Omega|t_j|^2$
from the background of the current noise  $\langle\!\langle I^2_j\rangle\!\rangle=\Omega|t_j|^2(1-|t_j|^2)$,
and thus, to measure the state of the system.
This measurement process is intimately related to dephasing: Since $|t_1t_2^*+r_1r_2^*|\leq 1$, off-diagonal
elements of the density matrix of the two-level system decay with the rate of the order of $\Omega$.

Next, we change the point of view and consider the two-level system as a detector of the current noise 
created by the QPC. The Ref.\ \onlinecite{levitov} proposes to place the two-level system to the right of the QPC and away
from the scattering region, so that the only affect of coupling is to induce the scattering phase shift $\lambda$, so that
$|t_1|^2=|t_2|^2={\cal T}$. Neglecting the overall phase shift, the function $h$ in Eq.\ (\ref{evolve}) acquires
the familiar form of the FCS generating function of the binomial process
\begin{equation}
h(\lambda)=\Omega\ln[1+{\cal T}(e^{i\lambda}-1)].
\label{generator}
\end{equation}
On one hand, the current cumulants can be obtained by differentiating the function $h(\lambda)$ and setting $\lambda=0$.
On the other hand, according to the Ref.\ \onlinecite{levitov} the physical effect of the noise on the two-level system 
arises because every
time an electron passes through the two-level system, it rotates the pseudo-spin by the phase $\lambda$.  
This happens randomly, according to the binomial statistics of transmissions of the QPC. Interestingly,
if $\lambda$ is not small, all the current cumulants contribute to this physical effect. On the contrary, if $\lambda\ll 1$,
which is a typical situation because it is difficult to arrange strong coupling, higher order current
cumulants have a negligible effect. In this situation, only Gaussian noise
can be directly detected by measuring the dephasing rate in the two-level system. This illustrates the 
central limit theorem at work: in the case of weak coupling, the system has to accumulate a large number
of small fluctuations
in order to experience a considerable change of state.

Remarkably similar situation arises in the case of a MZ interferometer exposed to the current noise of the QPC0 placed
upstream, despite the fact that this is a much more complex system. As discussed in the main part of the paper, 
current fluctuations randomly shift the AB phase of the 
interferometer. After averaging over these fluctuations, we have arrived at the visibility of AB oscillations,
which can be presented as $\nu={\rm Re}\rho_{12}$ with ${\cal T}$ replaced by ${\cal T}_0$.  
Therefore, by comparing to the two-level system
we can investigate the origin of the phase transition in the interferometer. We recall, that the phase transition
arises at ${\cal T}_0=1/2$ and $\lambda=\pi$. Setting in Eq.\ (\ref{state}) $|t_1|^2=|t_2|^2=1/2$ and 
the relative phase shift to $\pi$, we find that the two states $r_j|L\rangle+t_j|R\rangle$, $j=1,2$, are
mutually orthogonal, which implies the perfect entanglement between the two-level system and the reflected
electron. In this situation the measurement becomes projective and leads to complete dephasing, which explains
the divergence in $h(\lambda)$. We, therefore, conclude, that the phase transition in the MZ interferometer
is caused by the perfect entanglement between the electrons of the QPC0 and those of the interferometer.

%
%
%